# Piston sensing for sparse aperture systems via all-optical diffractive neural network

Xiafei Ma, Zongliang Xie, Haotong Ma, and Ge Ren

*Abstract*—It is a crucial issue to realize real-time piston correction in the area of sparse aperture imaging. This paper demonstrates that an optical diffractive neural network is capable of achieving light-speed piston sensing. By using detectable intensity distributions to represent pistons, the proposed method can convert the imaging optical field into estimated pistons without imaging acquisition and electrical processing, thus realizing the piston sensing task all-optically. The simulations verify the feasibility of the approach for fine phasing, with testing accuracy of λ/40 attained. This method can greatly improve the real-time performance of piston sensing and contribute to the development of sparse aperture system.

## I. INTRODUCTION

OPTICAL synthetic aperture imaging is an important technology approach to realize high-resolution imaging, exceeding the aperture size limits of the monolithic primary mirror systems with lighter weight and less manufacturing costs. However, the phase differences between sub-apertures adversely affect the optical interference and damage image quality. In order to achieve the optimal imaging performance, piston sensing technology, which acts a critical role in synthetic aperture imaging, is proposed to realize the co-phase between sub-apertures.

The present piston sensing methods can be categorized into two types, specific optics-based methods and image-based methods respectively. The modified Shack-Hartmann sensor [1], dispersed fringe sensor [2] and pyramid sensor [3] measure piston errors from the pupil information modulated by specially designed hardware, which inevitably increase the system complexity. The image-based methods, such as phase diversity [4] and phase retrieval [5], measure piston errors by analyzing the intensity distribution in image plane, for which a simpler optical design that requires significantly less hardware can produce the required image data. However, although the mentioned image-based piston sensing greatly compacts the sparse aperture system, it does need a large amount of iterative optimization calculation, thus failing to realize instant correction.

With the advent of era of Artificial Intelligence, neural network method has successfully resolved many optical imaging problems with piston sensing technology included [6-8]. Differing from the conventional optimization methods, the neural network method allows a computational model to holistically describe a specific problem by use of massive training data, thus making the prediction more efficient. It can be regarded that the offline time for iterative optimization is transferred to the online training stage. For the issue of piston sensing, the data-driven mode enables multi-layer network to extract abstract features by processing Point Spread Function (PSF), so as to establish the mapping relationship between image information and pistons. Guerra-Ramos et al. demonstrated the feasibility of deep learning-based piston sensing method in simulation. they trained two shallow convolutional neural networks respectively for fine phasing and coarse phasing, with multi-wavelength approach adopted to settle the 2π ambiguity problem and expand the capture range [9]. Subsequently, piston sensing method using a single network is developed, which is capable of predicting pistons directly from the raw broadband focal image via the execution of only one iteration in testing phase [10, 11]. This end-to-end mode makes the faster sensing realizable.

Benefited from the strong parallel computing capability of graphics processing unit (GPU), large amounts of mathematical operations in deep network and other iterative algorithms can be executed effectively. However, due to the inherent limitation of the von Neumann architecture in hardware processing speed, the traditional electrical neural network has encountered bottlenecks. Besides, the image acquisition speed of CCD will also hinder the real-time sensing of the image-based methods. Recently, diffractive optical elements (DOEs) have been used to build the deep learning framework and some achievements have been achieved in specific tasks, including image classification [12, 13], object detection [14], and pupil phase retrieval [15], among others. The parallel computing capability and high-speed optical transmission allow these tasks to be executed more efficiently.

Inspired by researches on diffractive network, we propose a

This paragraph of the first footnote will contain the date on which you submitted your paper for review, which is populated by IEEE. This work was supported in part by National Natural Science Foundation of China under Grant 62005289 and Grant 62175243; in part by Excellent Youth Foundation of Sichuan Scientific Committee under Grant 2019JDJQ0012; in part by Youth Innovation Promotion Association, CAS under Grant 2020372. in part by the National Key Research and Development Program of China under Grant 2022YFB3901900 and in part by Outstanding Scientist Project of Tianfu Qingcheng Program. *(Corresponding author: Xiafei Ma, Haotong Ma).*

The authors are with Key Laboratory of Optical Engineering, Institute of Optics and Electronics, Chinese Academy of Sciences, Chengdu, 610209, China (e-mail: maxiafei3860@163.com; mahaotong@163.com; renge@ioe.ac.cn, zongliang.xie@yahoo.com;).

real-time piston sensing method based on all-optical diffractive neural network (ODNN). The principle of the all-optical piston sensing is described in Section 2. The simulation results and some discussions are presented in Section3. By using a three-layer phase-only ONDD, a numerical blind testing accuracy of 97% is achieved for two-aperture imaging system, according to which a fine piston sensing with average RMSE of 0.024λ is obtained. In addition, it is also shown that a well-trained network has strong robustness for aberration. Finally, concluding thoughts and relevant discussions are offered in Section 4.

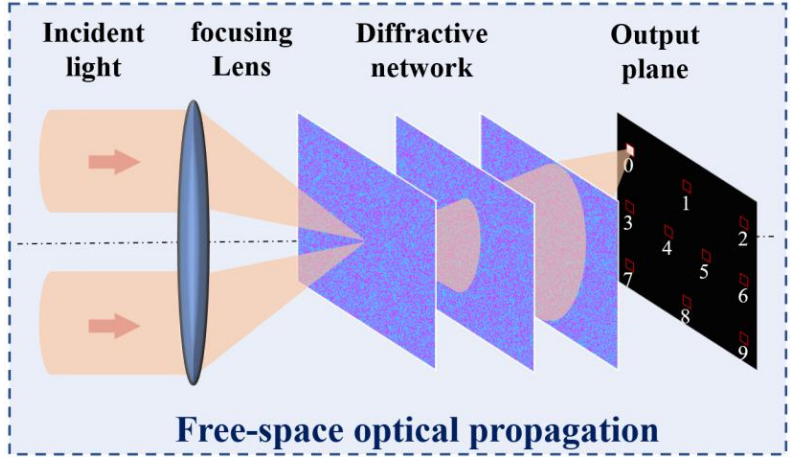

**Fig. 1.** Conceptual illustration of the piston sensing model using all-optical diffractive network.

## II. METHOD DESCRIPTION

Different from the electrical neural network method, where intensity images recorded by CCD are needed as the input data, ODNN composed of several DOEs is capable of processing the light field at the speed of light without any image acquisition and hardware processing. As conceptual illustration of the piston sensing model shown in Fig. 1, the incident light carrying piston information is focused on the first layer of the designed network, then the interference light field is modulated by the diffractive units and transformed to the specific intensity distribution, which corresponds to predicted piston value. The whole piston sensing process based on free-space optical propagation is implemented with all-optical computing, thus achieving light-speed sensing. It should be specially explained that there is no imaging device in the practical piston sensing module. For the purpose of high real-time sensing, several photosensitive elements will be placed at the output plane in the further experiment research, which will convert the optical signal into voltage signal, thus actuating the fast reflect mirror to compensate the piston errors.

The training flow diagram of the ODNN-based method is shown in Fig.2. To begin with, the simulation model of a sparse aperture imaging system is constructed to produce the complex amplitude of Point Spread Function (PSF), which is stored as two components: amplitude distribution and phase distribution. Then the diffractive network is trained to learn the mapping relationship between the complex-value PSF and intensity distribution at the output plane. Through iterative optimizations, optimum predicted intensity distributions are obtained with the convergence of loss function. In the computer-based training and testing, simulated complex-value PSF of sparse aperture system is generated and saved as input dataset. Once the network parameters are determined and the physical system is constructed, the digit input is no longer required.

According to the synthetic imaging principle, the generalized pupil function at the monochromatic wavelength λ can be expressed as:

$$A(x_0, y_0) = \sum_{n=1}^{N} A_{sub}(x_0 - x_n, y_0 - y_n) \exp(\frac{2\pi i}{\lambda} OPD_n). \quad (1)$$

where $(x_n, y_n)$ is the center vector of the n-th sub-aperture. $OPD_n$ is the piston term of the n-th sub-aperture relative to the reference. $A_{sub}(x_0 - x_1, y_0 - y_1)$ is defined as the reference sub-aperture with $OPD_1$ of zero. The complex amplitude on the focal plane of multi-aperture system, as referred as complex value PSF, can be calculated by implementing Fourier transform on the generalized pupil function:

$$U(x_f, y_f) = FT\{A(x_0, y_0)\}. \quad (2)$$

where FT(·) and $(x_f, y_f)$ represent Fourier transform and spatial coordinate on the focal plane.

For monochromatic source, the capture range (-0.5λ, 0.5λ) can be divided into 10 subranges at 0.1λ intervals, which are numbered P0-P9 {P0: (-0.5λ, -0.4λ), P1: (-0.4λ, -0.3λ), P2: (-0.3λ, -0.2λ), P3: (-0.2λ, -0.1λ), P4: (-0.1λ, 0), P5: (0, 0.1λ), P6: (0.1λ, 0.2λ), P7: (0.2λ, 0.3λ), P8: (0.3λ, 0.4λ), P9: (0.4λ, 0.5λ),}. Correspondingly, the output plane is divided into 10 subregions (R0-R9). For example, when the piston of system belongs to subrange P0: (-0.5λ, -0.4λ), the well-trained ODNN will ideally focus the maximum optical signal on the subregion 0 (R0), as the target shown in Fig. 2.

which constitute the ODNN is $M$, the intensity at the output plane (the $M+1$ layer) can be expressed as:

$$S_i^{M+1} = \left|u_i^{M+1}\right|^2. \qquad (6)$$

If the number of pixels at output plane is $K$, the loss function is defined as the mean squared error between the intensity on the output plane $S_k^{M+1}$ and the ideal intensity distributions $g_k^{M+1}$:

$$E = \frac{1}{K}\left(\sum_k S_k^{M+1} - g_k^{M+1}\right)^2, \qquad (7)$$

In the error backpropagation process, stochastic gradient descent algorithm is implemented to iteratively updating the network parameters until the goal of minimizing the loss function is attained. Then most of the output light is focused on the target region, thus achieving the mapping from imaging optical field to piston values. The final parameters of the model are saved and the architecture of the diffractive network used to perform piston sensing task is obtained. Once the model is physically fabricated, light-speed piston sensing task can be performed.

### III. RESULTS

To demonstrate the feasibility of the proposed ODNN-based piston sensing method, several simulations have been implemented. First, we model a two-aperture imaging system with 600 nm monochromatic light source, 10mm diameter sub-aperture, and 2m focal length according to the configuration of muti-aperture system in our lab. For each subrange, 1000 random values are generated as the pistons to produce input data of the network, which can be divided into 800 training samples and 200 testing samples. Then training dataset with 8000 samples and testing dataset with 2000 samples are obtained.

The ODNN utilized in this paper comprises three transmissive diffractive layers. Each neuron has a size of 15um and each layer composed of 160,000 neurons has a size of 6mm×6mm (400×400). The distance between the adjacent layers is 11cm and the first layer is located at the focal plane of the multi-aperture imaging system. The network in trained on a desktop computer with NVIDIA GTX 1080 Ti graphic card. The parameters are optimized by using AMSGrad algorithm [16] with a learning rate of 0.01 and batch size of 64. When the training stage is completed, the phase parameters of each diffractive layer are determined and diffractive network capable of executing piston sensing task all-optically is obtained. In the numerical simulations, the trained phase parameters are random values between 0 and $2\pi$. Considering the deviation in the manufacturing process, where eight-level binary optical technique will be utilized to fabricate the diffractive units, we finalize the phase values using eight-step quantization. All the simulation results in this paper are worked out with these quantized phase parameters.

Then simulation without aberration is investigated. In the numerical testing, testing dataset with 2000 samples are input to the ODNN and corresponding predicted intensity distributions are output. The confusion matrix for 1000 testing samples randomly selected from testing dataset is shown as Fig.

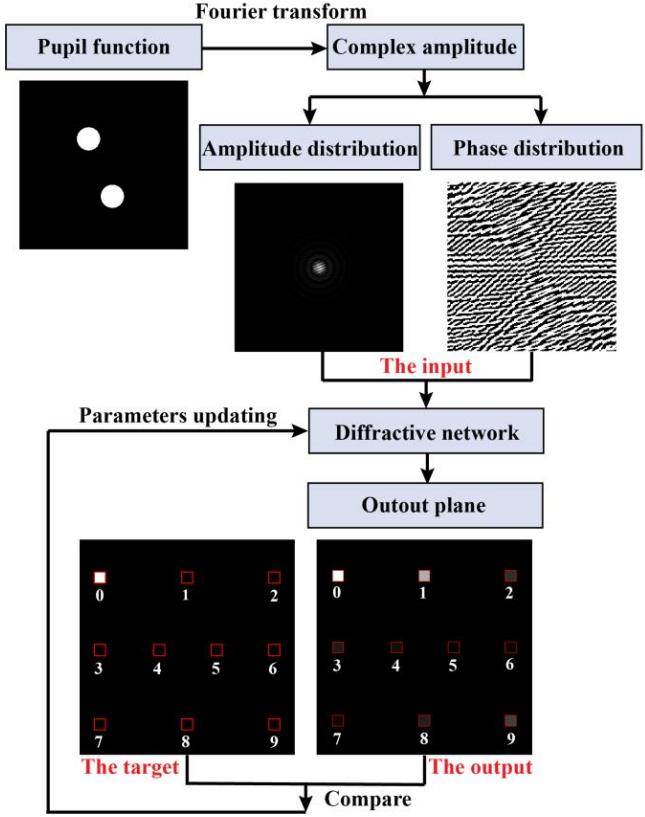

**Fig. 2.** Schematic of training procedure of the ODNN-based piston sensing method. The subregions numbered 0-9 (R0-R9) on the output plane correspond to 10 subranges (P0-P9) of the piston sensing range.

The information forward propagation process is realized through the diffraction between adjacent layers, the model of which can be established based on Rayleigh-Sommerfeld diffraction equation and written as:

$$w_i^l(x,y,z,l) = \frac{z-z_i}{r^2}\left(\frac{1}{2\pi r} + \frac{1}{j\lambda}\right)\exp\left(\frac{j2\pi r}{\lambda}\right), \qquad (3)$$

Where $r$ is the distance from the source to the neuron located at $(x_i, y_i, z_i)$ in the $l$-th layer. the transmission coefficients of the diffractive layers can be expressed as:

$$t_i^l(x_i, y_i, z_i) = a_i^l(x_i, y_i, z_i)\exp\left(j\varphi_i^l(x_i, y_i, z_i)\right), \qquad (4)$$

where $a_i^l(x_i, y_i, z_i)$ and $\varphi_i^l(x_i, y_i, z_i)$ represent the learnable amplitude parameter and phase parameter of each neuron. For the phase-only ODNN, the amplitude parameter is regarded as a constant equal to 1.

The output of the $i$-th neuron in the $l$-th layer is the product of its input complex amplitude information and transmission coefficients:

$$u_i^l(x,y,z) = w_i^l(x,y,z)t_i^l(x_i,y_i,z_i)\sum_k u_k^{l-1}(x_i,y_i,z_i), \qquad (5)$$

Where $\sum_k u_k^0(x_i, y_i, z_i)$ (when $l=1$) is the discretized complex-value PSF. When the number of diffractive layers

4(a), which describes the statistical overview of correct classification and incorrect classification for each subrange. According to the simulation results, the blind testing accuracy of the subrange prediction is approximately 97%.

Here we take the middle value of the predicted subrange as the piston sensing result. For example, when the predicted subrange is P9: (0.4λ, 0.5λ), the estimated piston value is regarded as 0.45λ. When the predicted labels are consistent with the ground-truth labels, the maximum deviation between the estimated result and the ground-truth piston value is 0.05λ. In the case of incorrect predictions, it can be seen that the wrong predicted labels are adjacent to the ground-truth label from the confusion matrixes in Fig. 4(a), which means the maximum deviation is no more than 0.15λ. The statistical result indicates that the average value of the RMSE for testing samples is about 0.024λ. The histograms in Fig. 3 display the intensity ratios of different subregions in 10 blind tests where the predicted labels are 0 to 9 respectively. it can be seen that there is a remarkable intensity contrast between the expected labels and the others. This is a meaningful index in the further piston correction stage, which can greatly prevents the possibility of misjudgment and assures the subrange prediction accuracy.

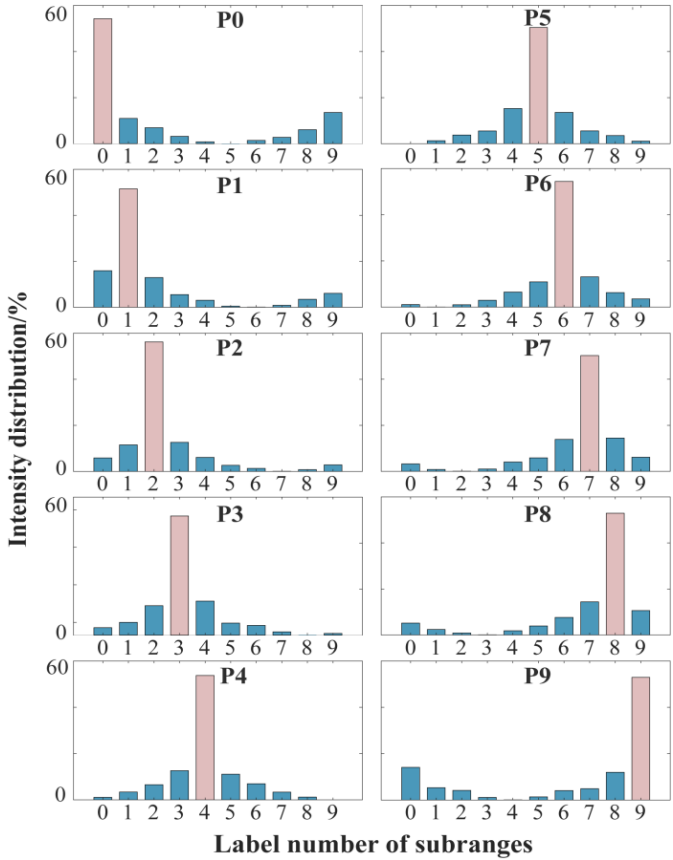

**Fig.3** Intensity distribution ratios of 10 subregions for imaging system without aberrations when the predicted labels are 0 to 9, respectively. The pink bar in each histogram corresponds to energy distribution percentage of the target subregion.

Then 0.05λ aberration generated by using 2-15 Zernike coefficients is introduced in the imaging system, and numerical simulation is implemented to evaluate the impact of aberration on this method. According to the confusion matrix shown in Fig. 4(b), the testing accuracy achieves 97% as well. When piston value is -0.18λ, which belongs to subrange 3, the intensity distributions at the output plane without aberration and with aberration are presented in Fig. 5(a) and Fig. 5(b) respectively. It is obvious that almost all the light energy is focused on the defined 10 subregions in both cases. Compared to the intensity distribution without aberration, where there is little background noise outside the subregions R0-R9, the noise on the undesirable regions is relatively greater when aberration exists. However, this background noise would not cause interference on the subregion recognition, which indicates that the proposed method has good robustness for aberration.

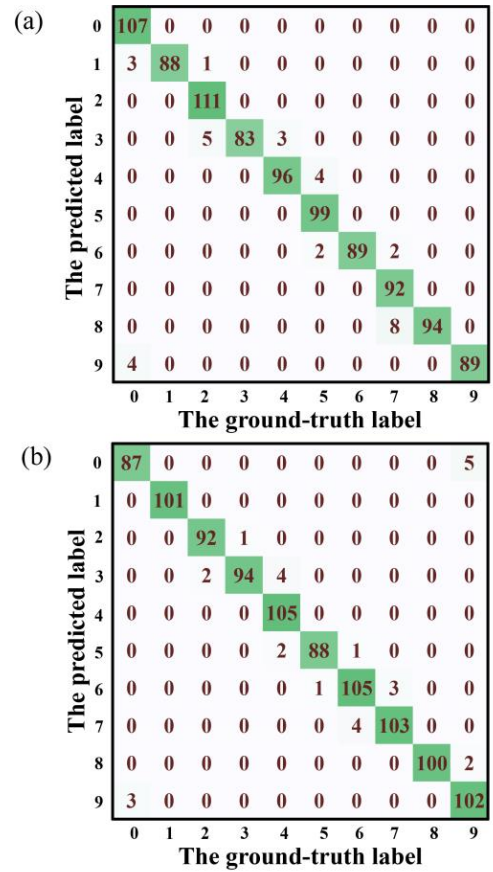

**Fig. 4.** (a) The confusion matrixes of testing results (a) without aberration and (b) with aberration.

There are some discussions about the design of the diffractive network for piston sensing task. The first one is about the neuron size. Larger neuron size can reduce the difficulties in fabrication as well as the alignment of network layers in experiment. However, for our multi-aperture imaging system, the PSF only occupies an area of hundreds of microns. Larger neuron will discrete the complex amplitude information into less pixels, and the resulting low sampling rate will lead to failure in network training. According to the hyper-parameter tuning results, neuron size is defined as 15um in our simulations. It should be noted that alignment error has a serious influence on the sensing accuracy. When alignment error is equal to one neuron size, the accuracy decreases from 97% to 54%. Thus translation stages with high precision is needed to control the alignment error within one neuron size. The second one is the axial distance between adjacent layers. On the one hand, the distance setting need to make the light beam cover more neurons in the forward-propagation, thus guaranteeing the higher-degree connectivity of the network. On the other hand, larger distance can improve the tolerance of the network to the alignment error in axial distance. As shown in Fig. 6, when the deviation between the successive layers is 1.25mm, the accuracies of the method decrease from 97% to 41% for distance of 4cm and 61% for distance of 8cm. Whereas, high accuracy of 92% is still achieved when the distance is 11cm. Reasonable parameters setting can effectively promote the piston sensing performance of the network.

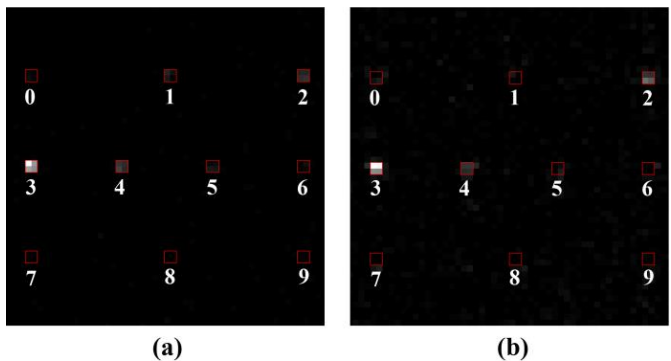

**Fig. 5.** The intensity distributions at the output plane for the simulated samples (a) without aberrations and (b) with aberrations.

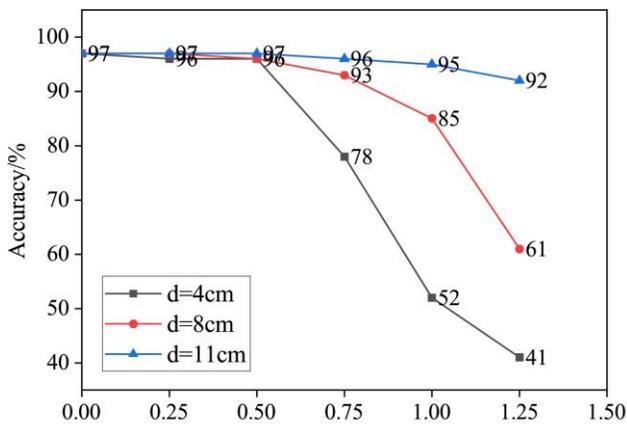

**Fig. 6.** The effect on accuracy of alignment error in axial distance with different distance parameters d.

V. CONCLUSION AND DISCUSSION

In conclusion, this paper has demonstrated that ODNN is capable of implementing piston sensing in an all-optical manner. The designed three-layer phase-only ODNN can directly convert the imaging optical field into intensity distribution which represents piston value. Fine phasing is achieved in the cases without aberration and with aberration. As a preliminary feasibility study, there are still some restrictions on the all-optical realization of piston sensing. Firstly, as categorization model is adopted instead of numerical fitting for the real-time performance, it becomes more intricate for the cases with more apertures. This may be resolved by using multiple diffractive networks to estimate piston value of every two sub-apertures. Additionally, network with smaller neurons and more compact design will be used in the further researches. Secondly, other superior piston representation models need to be explored to address the difficulties in piston sensing of extended objects.

REFERENCES

[1] G. Chanan, M. Troy, F. Dekens, S. Michaels, J. Nelson, T. Mast, and D. Kirkman, "Phasing the mirror segments of the Keck telescopes: the broadband phasing algorithm," *Appl. Opt.*, vol. 37, no. 1, pp. 140–155, 1998.
[2] M. A. van Dam, B. A. McLeod, and A. H. Bouchez, "Dispersed fringe sensor for the Giant Magellan Telescope," *Appl. Opt.*, vol. 55, no. 3, pp. 539-547, 2016.
[3] F. Shi, D. C. Redding, A. E. Lowman, C. W. Bowers, L. A. Burns, P. Petrone, C. M. Ohara, and S. A. Basinger, "Segmented mirror coarse phasing with a dispersed fringe sensor: experiment on NGST's wavefront control testbed," *Proc. SPIE*, vol. 4850, pp. 318–328, Aug. 2003.
[4] R. G. Paxman and J. R. Fienup, "Optical misalignment sensing and image reconstruction using phase diversity," *J. Opt. Soc. Am. A*, vol. 5, no. 6, pp. 914–923, 1988.
[5] D. S. Acton, J. S. Knight, A. Contos, S. Grimaldi, J. Terry, P. Lightsey, A. Barto, B. League, B. Dean, J. S. Smith, C. Bowers, D. Aronstein, L. Feinberg, W. Hayden, T. Comeau, R. Soummer, E. Elliott, M. Perrin, and C. W. Starr, "Wavefront sensing and controls for the James Webb space telescope," *Proc. SPIE*, vol. 8442, no. 84422H, 2012.
[6] A. Sinha, J. Lee, S. Li and G. Barbastathis, "Lensless computational imaging through deep learning," *Optica*, vol. 4, no. 9, pp. 1117-1125, 2017.
[7] L. Möckl, A. R. Roy and W. E. Moerner, "Deep learning in single-molecule microscopy: fundamentals, caveats, and recent developments," *Optica*, vol. 11, no.3, pp. 1633-1661, 2020.
[8] J. R. P. Angel, P. Wizinowich, M. Lloyd-Hart, and D. Sandler, "Adaptive optics for array telescopes using neural-network techniques," *Nature*, vol. 348, no. 6298, pp. 221–224, 1990.
[9] G. R. Dailos, D. G. Lara, T. S. Juan, and R. R. M. Jose, "Piston alignment of segmented optical mirrors via convolutional neural networks," *Opt. Lett.*, vol. 43, no. 17, pp. 4264-4267, 2018.
[10] X. Ma, Z. Xie, H. Ma, Y. Xu, G. Ren, and Y. Liu, "Piston sensing of sparse aperture systems with a single broadband image via deep learning," *Opt. Express*, vol. 27, no. 11, pp. 16058-16070, 2019.
[11] X. Ma, Z. Xie, H. Ma, Y. Xu, D. He, and G. Ren, "Piston sensing for sparse aperture systems with broadband extended objects via a single convolutional neural network," *Opt. Lasers Eng.*, Vol. 128, no. 106005, 2020.
[12] X. Lin, Y. Rivenson, N. T. Yardimci, M. Veli, Y. Luo, M. Jarrahi and A. Ozcan, "All-optical machine learning using diffractive deep neural networks," Science, vol. 361, no. 6406, pp. 1004-1008, 2018.
[13] H. Chen, J. Feng, M. Jiang, Y. Wang, J. Lin, J. Tan, and P. Jin, "Diffractive Deep Neural Networks at Visible Wavelengths," *Engineering*, vol. 7, no. 10, pp. 1483-1491, 2021.
[14] T. Zhou, X. Lin, J. Wu, Y. Chen, H. Xie, Y. Li, J. Fan, H. Wu, L. Fang and Q. Dai, "Large-scale neuromorphic optoelectronic computing with a reconfigurable diffractive processing unit," *Nature Photonics*, vol. 15, no. 5, pp. 367-373, 2021.
[15] X. Pan, H. Zuo, H. Bai, Z. Wu, and X. Cui, "Real-time wavefront correction using diffractive optical networks," Opt. Express, vol. 31, no. 2, pp. 1067-1078, 2023.
[16] S. J. Reddi, S. Kale and S. Kumar, "On the convergence of adam and beyond,", https://doi.org/10.48550/arXiv.1904.09237, 2018.